\pdfoutput=1

\documentclass[11pt,twoside,a4paper,cmspaper,final,collab]{cms-tdr}
\def\svnVersion{78710:78733}\def\cmsCernNo{2011-138}\def\cmsCernDate{\today}\def\cmsMessage{Submitted to Physical Review Letters}

\begin{document}\cmsNoteHeader{SUS-11-003}

\hyphenation{had-ron-i-za-tion}
\hyphenation{cal-or-i-me-ter}
\hyphenation{de-vices}
\RCS$HeadURL: svn+ssh://alverson@svn.cern.ch/reps/tdr2/papers/SUS-11-003/trunk/SUS-11-003.tex $
\RCS$Id: SUS-11-003.tex 78733 2011-09-11 20:30:44Z alverson $
\newlength\figwid\ifthenelse{\boolean{cms@external}}{\setlength\figwid{0.9\columnwidth}}{\setlength\figwid{0.8\textwidth}}
\newcommand{\scalht}{\mbox{$H_\text{T}$}\xspace}
\newcommand{\eslash}{{\hbox{$E$\kern-0.6em\lower-.05ex\hbox{/}\kern0.10em}}}
\newcommand{\met}{\mbox{$\eslash_\text{T}$}\xspace}
\newcommand{\cls}{\mbox{CL$_s$}\xspace}
\newcommand{\znunu}{\ensuremath{\cPZ \rightarrow \cPgn\cPagn}}
\newcommand{\wtaunu}{\ensuremath{\PW \rightarrow \Pgt\cPgn}}
\newcommand{\Et}{\ensuremath{{E_{\text T}}\xspace}}
\newcommand{\Hslash}{{\hbox{$H$\kern-0.8em\lower-.05ex\hbox{/}\kern0.10em}}}
\newcommand{\mht}{\mbox{$\Hslash_\text{T}$}\xspace}

\cmsNoteHeader{SUS-11-003}

\title{Search for Supersymmetry at the LHC in Events with Jets and
Missing Transverse Energy}

\date{\today}

\abstract{A search for events with jets and missing transverse energy
  is performed in a data sample of pp collisions collected at
  $\sqrt{s} = 7\TeV$ by the CMS experiment at the LHC. The analyzed
  data sample corresponds to an integrated luminosity of
  1.14~fb$^{-1}$. In this search, a kinematic variable, $\alpha_\text{T}$,
  is used as the main discriminator between events with genuine
  and misreconstructed missing transverse energy. No excess of events
  over the standard model expectation is found. Exclusion limits in
  the parameter space of the constrained minimal supersymmetric
  extension of the standard model are set. In this model, squark
  masses below $1.1\TeV$ are excluded at 95\% CL. Gluino masses below
  $1.1\TeV$ are also ruled out at 95\% CL for values of the universal
  scalar mass parameter below $500\GeV$.}

\hypersetup{ pdfauthor={CMS Collaboration},
             pdftitle={Search for Supersymmetry at the LHC in Events
             with Jets and Missing Transverse Energy},
             pdfsubject={CMS},
             pdfkeywords={CMS, physics, jets, missing energy, SUSY,
             Alpha_T}
             }

\maketitle

The standard model (SM) of particle physics is generally considered to
be valid only at low energy scales and is expected to be superseded by
a more complete theory at higher scales. Supersymmetric (SUSY)
extensions to the SM~\cite{ref:SUSY-1, ref:SUSY0, ref:SUSY1,
  ref:SUSY2, ref:SUSY3, ref:SUSY4, ref:hierarchy1, ref:hierarchy2}
introduce a large number of new particles with the same quantum
numbers as their SM partners, but differing by half a unit of spin. If
R-parity conservation~\cite{Farrar:1978xj} is assumed, supersymmetric
particles, such as squarks and gluinos, are produced in pairs and
decay to the lightest, stable supersymmetric particle (LSP). If the
LSP is neutral and weakly interacting, a typical signature is a final
state of multi-jets accompanied by significant missing transverse
energy, $\met$. Experiments at the Tevatron~\cite{CDFtrileptons,
  CDFLimits, D0Limits, Abazov200934},
$\mathrm{Sp\bar{p}S}$~\cite{UA1Limits, UA2Limits}, HERA~\cite{HERALimits1,
  HERALimits2}, and LEP~\cite{LEPLimits} colliders have performed
extensive searches for signs of SUSY. In 2010, the Large Hadron
Collider (LHC) at CERN delivered an integrated luminosity of almost
50~pb$^{-1}$ at a centre-of-mass energy $\sqrt{s} = 7\TeV$, leading to
several new searches from both ATLAS~\cite{PhysRevLett.106.131802,
  Aad2011186, Aad2011398, springerlink:10.1140/epjc/s10052-011-1682-6}
and CMS~\cite{RA1Paper, RA2, Razor, PhysRevLett.106.211802,
  springerlink:10.1007/JHEP07(2011)113}.

This Letter presents a search for SUSY based on a data sample
corresponding to an integrated luminosity of $1.14 \pm
0.05$~fb$^{-1}$. The search strategy follows Ref.~\cite{RA1Paper} and
is designed to be sensitive to \met signatures in events with two or
more energetic jets. The search is not optimized for any particular
model of SUSY and is applicable to other new physics scenarios with a
\met signature. In this Letter, nevertheless, the results are
interpreted in the constrained minimal supersymmetric extension of the
standard model (CMSSM)~\cite{ref:MSUGRA, PhysRevLett.69.725,
  ref:CMSSM}.  The CMSSM is described by the following five
parameters: the universal scalar and gaugino mass parameters, $m_0$
and $m_{1/2}$; the universal trilinear soft SUSY-breaking parameter,
$A_0$; the ratio of the vacuum expectation values of the two Higgs
doublets, $\tan\beta$; and the sign of the Higgs mixing parameter,
$\mu$. We consider only parameter sets for which the LSP is the
lightest neutralino. The following example parameter set, referred to
as LM6, is used to illustrate possible CMSSM yields: $m_0 = 85\GeV$,
$m_{1/2} = 400\GeV$, $A_0 = 0$, $\tan\beta = 10$, and $\mu > 0$.

A detailed description of the CMS apparatus can be found in
Ref.~\cite{ref:CMS}. Its central feature is a superconducting solenoid
providing an axial magnetic field of 3.8~T. The bore of the solenoid
is instrumented with several particle detection systems.  Charged
particle trajectories are measured by a silicon pixel and strip
tracker system, with full azimuth ($\phi$) coverage and a
pseudorapidity ($\eta$) acceptance from $-2.5$ to $+2.5$. Here, $\eta
\equiv -\ln [ \tan (\theta/2) ]$ and $\theta$ is the polar angle with
respect to the counterclockwise beam direction. A lead tungstate
crystal electromagnetic calorimeter (ECAL) and a brass/scintillator
hadron calorimeter surround the tracking volume and provide coverage
in $\eta$ from $-3$ to $+3$. The forward hadron calorimeter extends
symmetrically the coverage by a further two units in $\eta$. Muons are
identified in gas ionization detectors embedded in the steel return
yoke of the magnet. The CMS detector is nearly hermetic, which allows
for momentum-balance measurements in the plane transverse to the beam
axis.

The offline event reconstruction and selection criteria described
below are explained in more detail in Ref.~\cite{RA1Paper}. Jets are
reconstructed from the energy deposits in the calorimeter towers,
clustered by the anti-$k_{\rm T}$ algorithm~\cite{antikt} with a size
parameter of $0.5$. The raw jet energies measured by the calorimeter
systems are corrected to establish a uniform relative response in
$\eta$ and a calibrated absolute response in transverse momentum \pt
with an associated uncertainty between $2\%$ and $4\%$, depending on
the jet $\eta$ and \pt~\cite{2011arXiv1107.4277C}. Jets considered in
the analysis are required to have transverse energy $\Et >
50\GeV$. Events are vetoed if any additional jet satisfies $\Et >
50\GeV$ and $|\eta| > 3$, or rare, spurious signals are identified in
the calorimeters~\cite{1748-0221-5-03-T03014, CMS-NOTE-2010-012}. The
highest-$\Et$ jet is required to be within the central tracker
acceptance and the two highest-$\Et$ jets must each have $\Et >
100\GeV$. To suppress SM processes with genuine \met from neutrinos,
events containing an isolated electron~\cite{PAS-EGM-10-004} or
muon~\cite{PAS-MUO-10-002} with $\pt > 10\GeV$ are vetoed. To select a
pure multi-jet topology, events are vetoed in which an isolated
photon~\cite{PAS-EGM-10-006} with $\pt > 25\GeV$ is found.

The following two variables characterize the visible energy and
missing momentum in the transverse plane: the scalar sum of the
transverse energy $\Et$ of jets, defined as $\scalht =
\sum_{i=1}^{N_{\rm jet}} \Et$, and the magnitude of the vector sum of
the transverse momenta $\vec{\pt}$ of jets, defined as $\mht =
|\sum_{i=1}^{N_{\rm jet}} \vec{\pt}|$, where $N_{\rm jet}$ is the
number of jets with $\Et > 50\GeV$. Significant hadronic activity in
the event is ensured by requiring $\scalht > 275\GeV$.  Following
these selections, the background from multi-jet production, a
manifestation of quantum chromodynamics (QCD), is still several orders
of magnitude larger than the typical signal expected from SUSY. While
the bulk of these multi-jet events do not exhibit significant \met,
large values can be observed due to stochastic fluctuations in the
measurement of jet energies or mismeasurements caused by
nonuniformities in the calibration of the calorimeters or detector
inefficiencies.

The $\alpha_{\text{T}}$ kinematic variable, first introduced in
Refs.~\cite{Randall:2008rw, cms-pas-sus-08005, cms-pas-sus-09001}, is
used in the selection to efficiently reject events either without
significant \met or with transverse energy mismeasurements, while
retaining a large sensitivity to new physics with genuine \met
signatures. For events with two jets, the variable is defined as
$\alpha_{\text{T}} = \Et^{\rm j_2}/M_\text{T} = \Et^{\rm
  j_2}/\sqrt{\scalht^2 - \mht^2}$, where $\Et^{\rm j_2}$ is the
transverse energy of the less-energetic jet, and $M_\text{T}$ is the
transverse mass of the dijet system. For a perfectly measured dijet
event with $\Et^{\rm j_1} = \Et^{\rm j_2}$ and jets back to back in
$\phi$, and in the limit of large jet momenta compared to their
masses, the value of $\alpha_{\text{T}}$ is 0.5. In the case of an
imbalance in the measured transverse energies of back-to-back jets,
$\alpha_{\text{T}}$ is smaller than 0.5. Values significantly greater
than 0.5 are observed when the two jets are not back to back and
balancing genuine \met. For events with three or more jets, a dijet
system is formed by combining the jets in the event into two {\it
  pseudo-jets}. The total $\Et$ of each of the two pseudo-jets is
calculated as the scalar sum of the measured $\Et$ of contributing
jets.  The combination chosen is the one that minimizes the $\Et$
difference between the two pseudo-jets. This simple clustering
criterion provides the best separation between QCD multi-jet events
and events with genuine \met. Events with multiple jets with $\Et <
50\GeV$ or with severe jet energy under-measurements due to detector
inefficiencies can lead to values of $\alpha_{\rm T}$ slightly above
0.5. Such events are effectively rejected by requiring
$\alpha_{\text{T}} > 0.55$ and by applying dedicated vetoes, described
further in Ref.~\cite{RA1Paper}. These final selections complete the
definition of the hadronic signal sample. A disjoint hadronic control
sample consisting predominantly of QCD multi-jet events is defined by
requiring $\alpha_{\text{T}} < 0.55$.

\begin{figure}[bht]
  \begin{center} \includegraphics[width = \figwid]{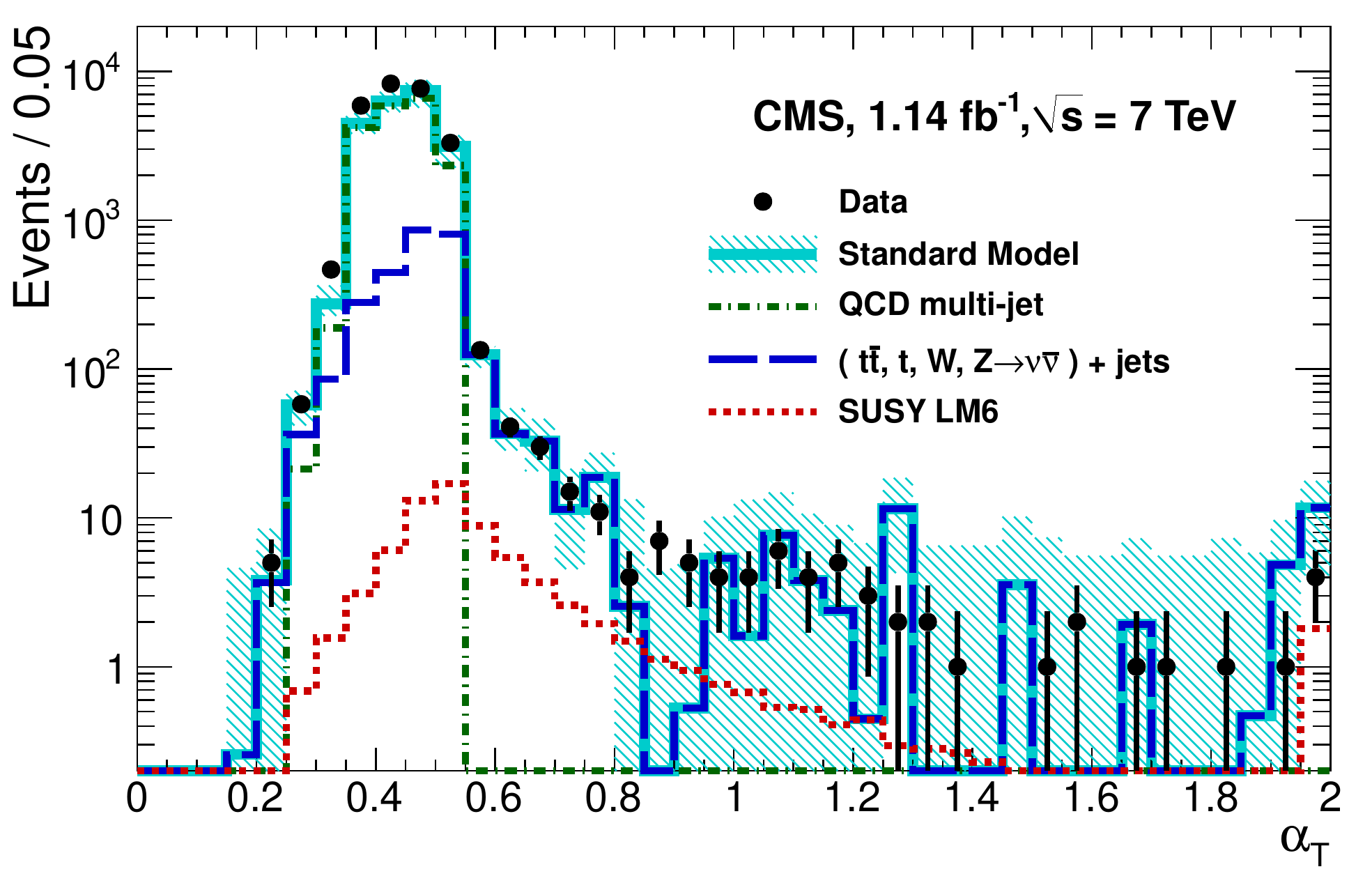}
    \caption{\label{fig:aT} The distribution of $\alpha_{\text{T}}$,
      described in the text, for events in data with two or more jets
      (black dots with error bars representing the statistical
      uncertainties), after all event selection criteria except
      $\alpha_{\text{T}}$ are applied and $\scalht > 375\GeV$. For
      illustrative purposes only, expected yields from simulation are
      also shown for QCD multi-jet events (dotted-dashed line),
      associated production of top quarks, W, or Z with jets
      (long-dashed line), the sum of all aforementioned SM processes
      (solid line) and the SUSY LM6 model (dotted line). The
      uncertainties for the SM expectation, due to the limited
      accuracy of the available simulation datasets and jet energy
      calibrations, are represented by the hatched area. The highest
      bin contains the overflows.}
    \end{center}
\end{figure}

As can be seen in Fig.\,\ref{fig:aT}, the only expected remaining
backgrounds with $\alpha_{\text{T}} > 0.55$ stem from SM processes
with genuine \met in the final state. In the dijet case, the largest
backgrounds with genuine \met are the associated production of W or Z
bosons with jets, followed by either the weak decays \znunu\ or
\wtaunu, where the $\tau$ decays hadronically and is identified as a
jet; or by leptonic decays that are not rejected by the dedicated
electron or muon vetoes. At higher jet multiplicities, top quark
production, followed by semileptonic weak top quark decay, becomes
important.

Events in the hadronic signal sample are recorded with a trigger
condition that identifies candidate events with energetic jets and
significant \met. Events are selected if they have $\scalht > 250\GeV$
and $\mht$ above a threshold that evolves with instantaneous
luminosity, from 60 to $90\GeV$. In the region $275 < \scalht <
325\GeV$, the efficiency with which events satisfying the full
reconstruction and selection criteria are triggered is
$0.99^{+0.01}_{-0.02}$. For events with $\scalht > 325\GeV$, the
efficiency is $1.00^{+0.00}_{-0.03}$. A set of pre-scaled $\scalht$
trigger conditions are used to record events for the hadronic control
sample.

The analysis makes use of two additional data samples to estimate the
backgrounds with genuine \met. First, a $\mu$ + jets sample is
recorded with the hadronic trigger condition described above. The
event selection, following closely the prescription described in
Ref.~\cite{toppaper}, requires a single, isolated muon with $\pt >
10\GeV$ in the final state and the transverse mass of the muon and
$\mht$ system to be larger than $30\GeV$ to ensure a sample rich in W
bosons. The muon is required to be separated from the closest jet in
the event by $\Delta \eta$ and $\Delta \phi$ such that the distance
$\Delta R \equiv \sqrt{\Delta \eta^2 + \Delta \phi^2} > 0.5$. Second,
a $\gamma$ + jets sample is selected using a dedicated photon trigger
condition requiring a localized, large energy deposit in the ECAL with
$\pt > 90\GeV$ and that satisfies loose photon identification and
isolation criteria~\cite{PAS-EGM-10-006}. The offline selection
requires a single photon to be reconstructed with $\pt > 100\GeV$,
$|\eta| < 1.45$, satisfying tight isolation criteria, and with a
minimum distance to any jet of $\Delta R > 1.0$.  For these selection
criteria, the photon trigger condition is found to be fully efficient.

The hadronic signal region is divided into eight bins of $\scalht$:
two bins of width $50\GeV$ in the range $275 < \scalht < 375\GeV$,
five bins of width $100\GeV$ in the range $375 < \scalht < 875\GeV$,
and a final open bin, $\scalht > 875\GeV$. As in Ref.~\cite{RA1Paper},
jet $\Et$ thresholds are scaled down from their nominal values in the
lowest two \scalht bins to maintain jet multiplicities and thus
comparable event kinematics, topologies, and background composition
throughout the entire \scalht range. The background estimation methods
described below are combined in the statistical interpretation of the
observed data yields to provide a single prediction of the SM
background in each \scalht bin of the hadronic signal region. With
respect to Ref.~\cite{RA1Paper}, these refinements provide greater
sensitivity across a broader SUSY parameter space and, in the context
of CMSSM, up to higher-mass states.

The $\mu$ + jets data sample provides an estimate of the contributions
from top quark and W production (leading to W + jets final states)
still remaining in each $\scalht$ bin of the hadronic signal region
after all selection criteria are applied. Factors obtained from
simulation~\cite{RA1Paper} are then used to translate the yields in
the $\mu$ + jets sample to estimates in each \scalht bin of the
hadronic signal region. These factors are found to be only weakly
dependent on \scalht, ranging from 1.14 at low \scalht to 0.90 at high
\scalht. Conservative uncertainties on all the parameters entering
these translation factors are assigned. The total systematic
uncertainty is estimated to be 30\%, dominated by the uncertainty on
the efficiency for vetoing leptons. The remaining irreducible
background of \znunu\ + jets events in the hadronic signal sample is
estimated from $\gamma$ + jets events. These two processes have
similar kinematic properties when the photon is
ignored~\cite{PAS-SUS-08-002, Bern:2011pa}, while the latter has a
larger production cross section. Translation factors that account for
the ratio of cross sections for $\gamma$ + jets and \znunu\ + jets,
and their relative acceptances, are obtained from
simulation~\cite{RA1Paper} and used to estimate the number of \znunu\
+ jets events in each \scalht bin of the hadronic signal region. As is
the case of the $\mu$ + jets sample, these translation factors are
only weakly dependent on \scalht, ranging from 0.35 at low \scalht to
0.45 at high \scalht.  The main systematic uncertainties on these
factors are associated with the ratio of cross sections between
$\gamma$ + jets and \znunu\ + jets in the simulation (30\%), the
efficiency for photon identification (20\%), and the purity of the
photon selection (20\%), which add up in quadrature to 40\%.

Furthermore, the $\scalht$ dependence of the ratio
$R_{\alpha_\text{T}}$ is exploited to constrain the SM background
estimate for each $\scalht$ bin. This ratio is defined as the number
of events with $\alpha_{\text{T}}$ above and below a threshold value
of 0.55 for a given bin in $\scalht$. The denominator of the ratio is
always dominated by events from QCD multi-jet production and is
measured in data with samples selected by the set of pre-scaled
\scalht trigger conditions. The chosen $\alpha_{\text{T}}$ threshold
ensures that, for a given bin in \scalht, the numerator of the ratio
is dominated by events from SM processes with genuine \met from
neutrinos, with no significant contribution from QCD multi-jet
production. As observed in Ref.~\cite{RA1Paper}, this property leads
to $R_{\alpha_\text{T}}$ being independent of $\scalht$. The remaining
backgrounds are those with genuine \met from associated production of
top quarks, W, or Z with jets. By relaxing the $\alpha_{\text{T}}$
threshold to values lower than 0.55, the numerator is instead
dominated by mismeasured QCD multi-jet events, and an exponential
dependence of $R_{\alpha_{\rm{T}}}$ on \scalht is
observed~\cite{RA1Paper}. The behaviours of $\alpha_{\rm{T}}$ and
$R_{\alpha_\text{T}}$($\scalht$) are observed in data and simulation
to be robust against the effects of multiple pp collisions per beam
crossing ({\it pileup}). In the statistical interpretation of the
analysis, $R_{\alpha_\text{T}}$($\scalht$) is modelled as a
superposition of a \scalht-independent contribution from SM processes
with genuine \met, and an exponentially falling contribution to
accommodate any potential QCD contamination. The latter is considered
even though no evidence of a significant QCD contamination is found in
the hadronic signal region.

To obtain an accurate and consistent prediction of the SM background,
a simultaneous binned likelihood fit using information from all three
data samples is performed. The fit maximizes the likelihood $L_\text{
total} = L_\text{hadronic} \times L_{\mu+\text{jets}} \times L_{\gamma
+ \text{jets}}$, where $L_\text{hadronic}$ characterizes
$R_{\alpha_{\rm{T}}}$($\scalht$) in the hadronic sample with a single
exponential function, $A\mathrm{e}^{-k \, H_\text{T}}$, to accommodate any
QCD contamination and a constant, \scalht-independent contribution,
$B$, to describe SM processes with genuine \met. The likelihoods
$L_{\mu + \text{jets}}$ and $L_{\gamma+ \text{jets}}$ describe the
\scalht-dependent yields in the $\mu$ + jets and $\gamma$ + jets
samples.  For each \scalht bin, these yields are related to the
numerator of $R_{\alpha_\text{T}}$($\scalht$), measured in the
hadronic signal sample, via the translation factors from the
simulation.

\begin{figure}[htbp]
  \begin{center}
    \includegraphics[width = \figwid]{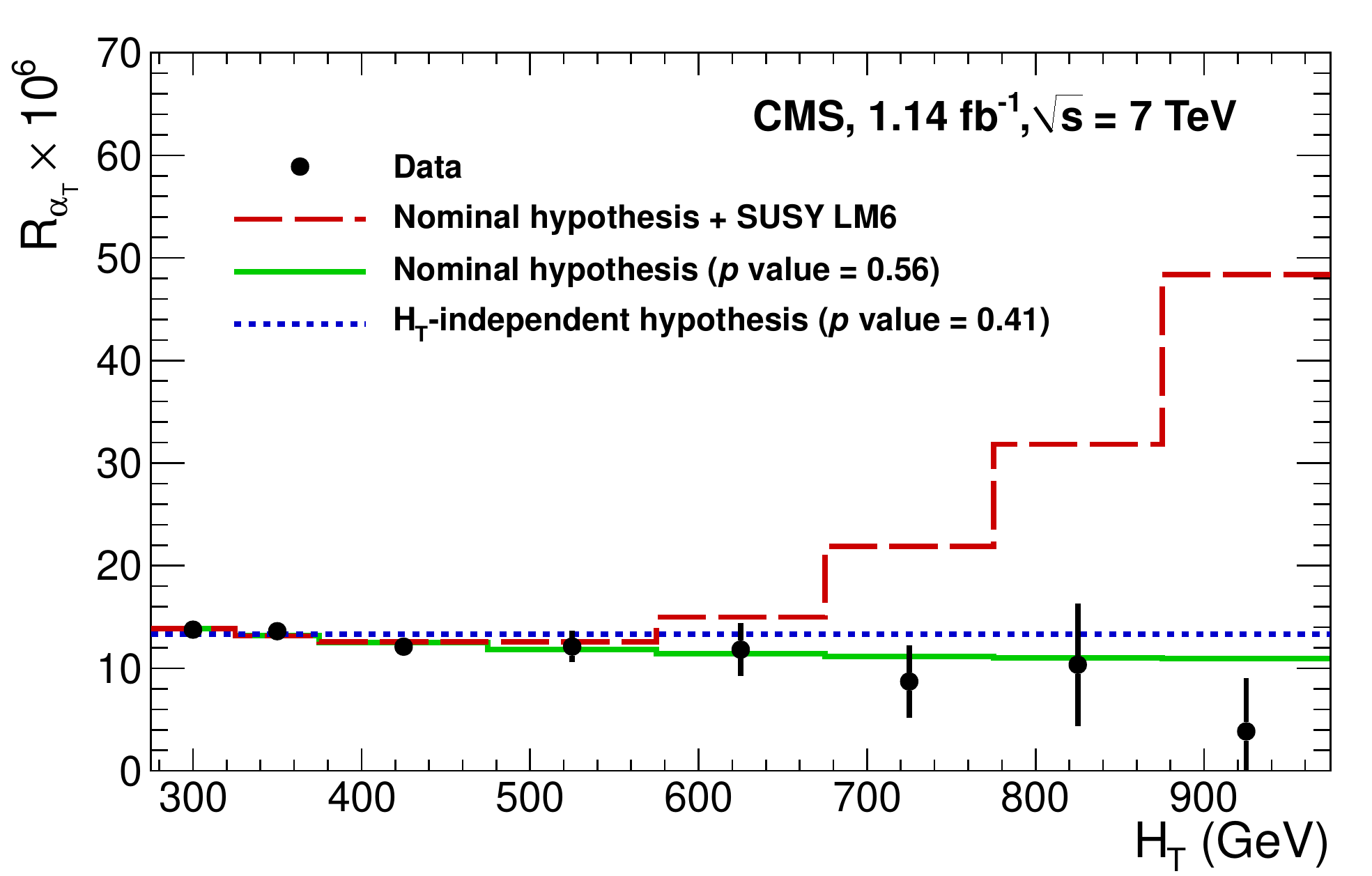}
    \caption{\label{fig:rat} The ratio $R_{\alpha_\text{T}}$ as a
      function of \scalht, as measured in the hadronic data samples
      (black dots with error bars representing the statistical
      uncertainties). Also shown is the result of the simultaneous fit
      to the three data samples (solid line); the analogous result
      when assuming a \scalht-independent hypothesis (dotted line);
      and, for illustrative purposes only, the expectation from the
      SUSY LM6 model when superimposed on the nominal fit result
      (long-dashed line).}
  \end{center}
\end{figure}

With a fit probability ($p$~value) of 0.56, the hypothesis for the
$R_{\alpha_\text{T}}$ dependence on \scalht reproduces the data well,
as shown in Fig.\,\ref{fig:rat}. The only parameter with a significant
non-zero value as determined by the fit is the constant term $B=(1.1
\pm 0.2) \times 10^{-5}$. The other two fit parameters, $A=(1.4 \pm
1.9) \times 10^{-5}$ and $k=(5.2 \pm 5.6)\times 10^{-3} \GeV^{-1}$,
are compatible with zero, indicating that no significant QCD
contamination has been observed in the signal region. Furthermore, as
a cross check, the fit is repeated with the assumption that
$R_{\alpha_\text{T}}$ is independent of \scalht, which in turn implies
that the numerator of the ratio is fully dominated by SM backgrounds
with genuine \met. The result of this fit, shown in
Fig.\,\ref{fig:rat}, has a $p$~value of 0.41 and is in good agreement
with the nominal fit.

\begin{table*}[htbp]
  \caption{Comparison of the measured yields in the different
    \scalht bins for the hadronic, $\mu$ + jets and $\gamma$ + jets
    samples with the SM expectations and combined statistical and
    systematic uncertainties given by the
    simultaneous fit.}
  \label{tab:results-fit}
  \centering \footnotesize
   \begin{tabular}{ ccccccccc }
     \hline
     \scalht bin (GeV)            & 275--325 & 325--375 & 375--475 & 475--575   & 575--675 & 675--775 & 775--875 & $>$875 \\ [0.5ex]
     \hline
     SM hadronic & $787^{+32}_{-22}$ &  $310^{+8}_{-12}$  &  $202^{+9}_{-9}$    &  $60.4^{+4.2}_{-3.0}$  &  $20.3^{+1.8}_{-1.1}$ &   $7.7^{+0.8}_{-0.5}$   &   $3.2^{+0.4}_{-0.2}$    &   $2.8^{+0.4}_{-0.2}$      \\
      Data hadronic   & 782 & 321& 196 & 62 & 21&  6 & 3 & 1         \\  \hline
     SM $\mu$ + jets                & $367^{+15}_{-15}$ &  $182^{+8}_{-9}$   &  $113^{+8}_{-7}$   &   $36.5^{+3.8}_{-3.3}$ &  $13.4^{+2.2}_{-1.8}$    &   $4.0^{+1.4}_{-1.2}$     &   $0.8^{+0.9}_{-0.1}$     &   $0.7^{+0.9}_{-0.1}$        \\
     Data $\mu$ + jets            & 389 & 156 & 113 & 39 & 17 &  5 & 0 & 0  \\ \hline
     SM $\gamma$ + jets   &  $834^{+28}_{-30}$ &  $325^{+17}_{-17}$  &  $210^{+12}_{-12}$    &  $64.7^{+6.9}_{-7.0}$  &  $21.1^{+3.9}_{-4.3}$ &   $10.5^{+2.5}_{-2.6}$   &   $6.1^{+0.9}_{-1.7}$    &   $5.5^{+0.9}_{-1.6}$        \\
      Data $\gamma$ + jets       &  849 &  307 & 210 & 67 & 24 & 12 & 4 & 4                       \\
     \hline
     \end{tabular}
\end{table*}

\begin{figure}[htbp]
  \begin{center}
    \includegraphics[width = \figwid]{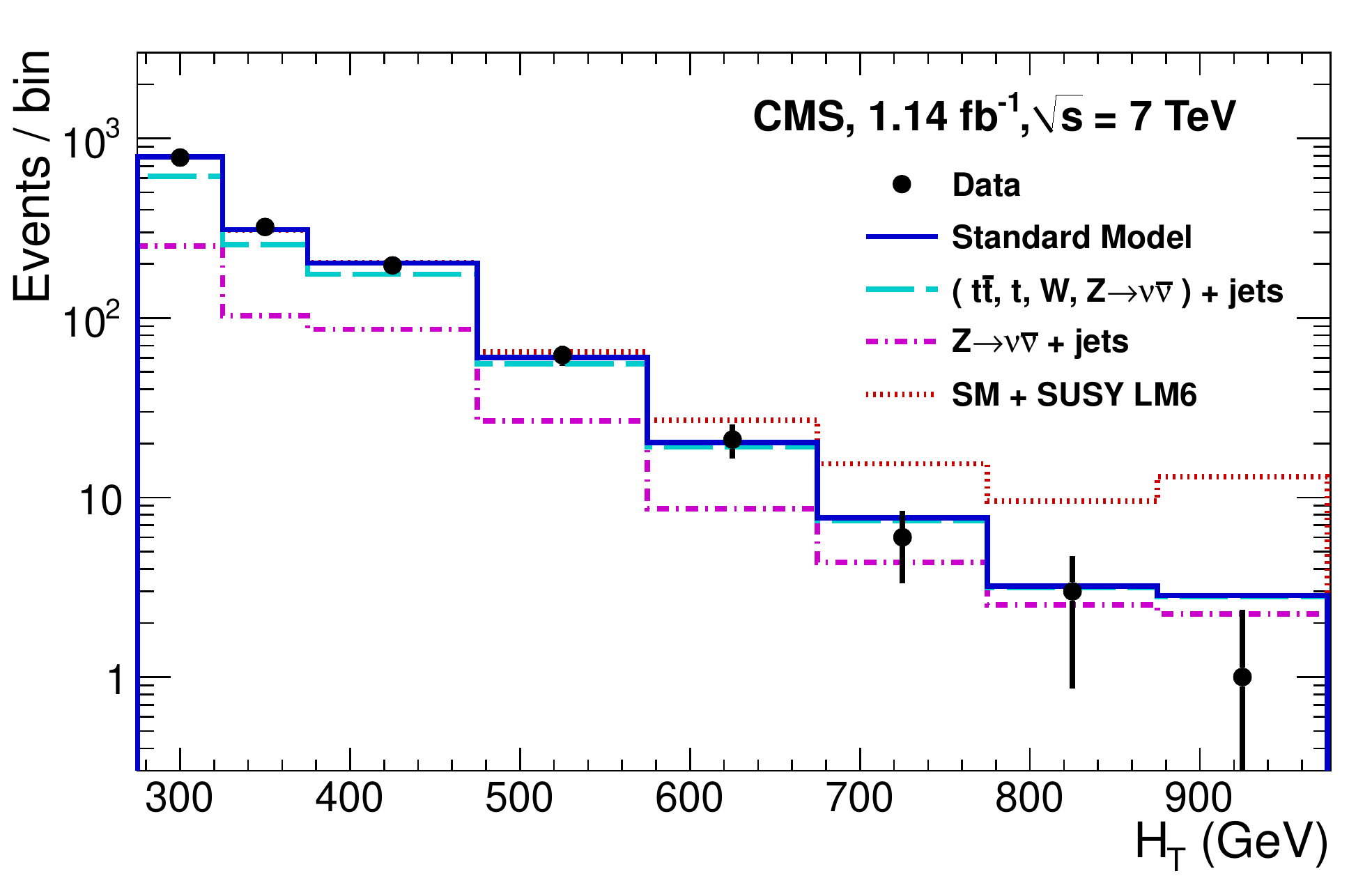}
    \caption{\label{fig:hadronic} The observed event yields versus
      \scalht in the hadronic signal sample (black dots with error
      bars representing the statistical uncertainties). Also shown are
      the expectations given by the simultaneous fit for the
      \znunu\ + jets process (dotted-dashed line);
      the associated production of top, W, or Z with jets (long-dashed
      line); the sum of QCD and all aforementioned processes (solid
      line); and, for illustrative purposes only, the SUSY LM6 model
      superimposed on the SM expectation (dotted line).}
  \end{center}
\end{figure}

The fit results for all three data samples are listed in
Table~\ref{tab:results-fit}, along with the observed yields in the
data. Good agreement between the measured \scalht distribution and the
fit is observed for all three data samples, indicating that the
observed yields are compatible with the SM background expectation
provided by the fit. The uncertainties listed with the SM predictions
are obtained from an ensemble of pseudo-experiments.
Figure\,\ref{fig:hadronic} compares the result of the simultaneous fit
to the observed yields in the hadronic signal sample.

Given the lack of an excess of events above the expected SM
backgrounds, limits are set in the parameter space of the CMSSM. At
each point of the parameter space, the SUSY particle spectrum is
calculated with \textsc{SoftSUSY} \cite{Allanach:2001kg}, and the signal
events are generated at leading order with
\PYTHIA6.4~\cite{pythia}. Inclusive, process-dependent,
next-to-leading-order (NLO) cross sections, obtained with the program
\textsc{Prospino}~\cite{Beenakker:1996ch}, are used to calculate the
observed and expected exclusion contours. The simulated signal events
are reweighted so that the distribution of number of pileup events per
beam crossing from the simulation matches that observed in
data. Uncertainties on the SM background prediction, the luminosity
measurement (4.5\%)~\cite{ref:lumi}, the total acceptance times
efficiency of the selection for the considered signal model
(4.5\%)~\cite{RA1Paper, PAS-JME-10-010}, and NLO cross section and
parton distribution functions (10\%) are included in the
limit. Although signal contributions to the total yield in each of the
three considered data samples are allowed, the only relevant signal
contribution originates from the hadronic data sample in the case of
the CMSSM.

\begin{figure}[htb]
  \begin{center}
    \includegraphics[width = \figwid]{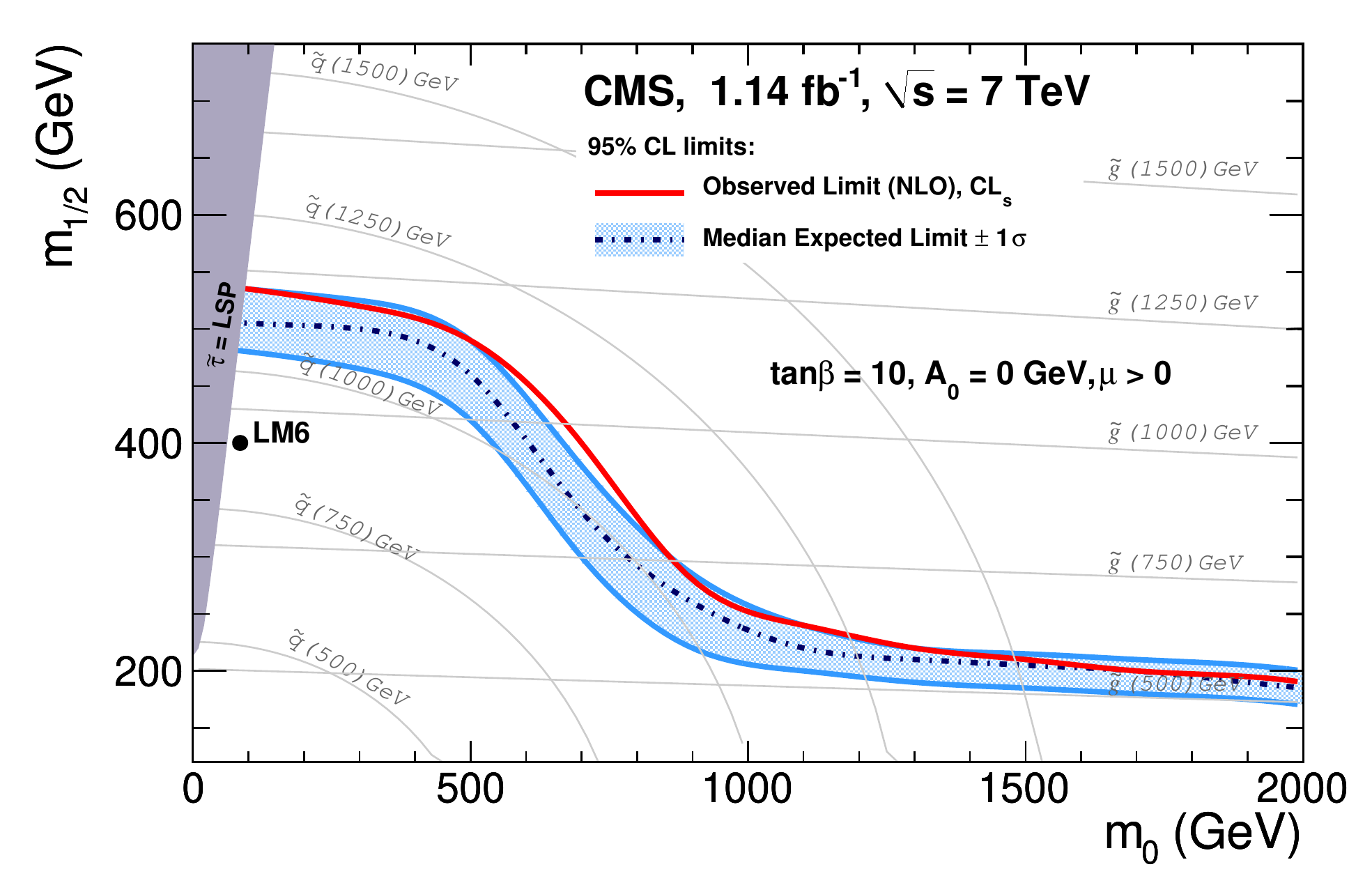}
    \caption{\label{fig:cmssm} Observed and expected 95\% CL exclusion
    contours in the CMSSM ($m_0, m_{1/2}$) plane ($\tan \beta = 10,
      A_0 = 0, \mu > 0$) using NLO signal cross sections with the \cls
      method. The expected limit is shown with its 68\% CL range. The
      SUSY benchmark model LM6 is also shown.}
  \end{center}
\end{figure}

Figure\,\ref{fig:cmssm} shows the observed and expected exclusion
limits at 95\% confidence level (CL) in the $(m_0 , m_{1/2})$ plane
for $\tan \beta = 10$ and $A_0 = 0\GeV$, calculated with the \cls
method~\cite{cls-pdg}. For this choice of parameters in the CMSSM,
squark masses below $1.1\TeV$ are excluded and gluino masses below the
same value are ruled out for $m_0 < 500\GeV$. The exclusion limit
changes at most by $20\GeV$ in the $(m_0 , m_{1/2})$ plane for
different parameter values (\eg $\tan \beta = 40$ and $A_0 =
-500\GeV$), indicating that the limit is only weakly dependent on
these parameters.

In summary, the first search for supersymmetry from CMS based on an
integrated luminosity in excess of 1~fb$^{-1}$ has been
reported. Final states with two or more jets and significant \met, as
expected from high-mass squark and gluino production and decays, have
been analysed. The search has been performed in eight bins of the
scalar sum of the transverse energy of jets, \scalht, considering
events with $\scalht$ in excess of $275\GeV$. The sum of standard
model backgrounds per $\scalht$ bin has been estimated from a
simultaneous binned likelihood fit to hadronic, $\mu$ + jets, and
$\gamma$ + jets samples. The observed yields in the eight \scalht bins
have been found to be in agreement with the expected contributions
from standard model processes. Limits on the CMSSM parameters have
been derived and squark masses below $1.1\TeV$ are excluded at 95\% CL
in this model. Gluino masses in the same range are ruled out at 95\%
CL for $m_0 < 500\GeV$. This limit represents a tight constraint on
the parameter space of SUSY models like the CMSSM.

We wish to congratulate our colleagues in the CERN accelerator
departments for the excellent performance of the LHC machine. We thank
the technical and administrative staff at CERN and other CMS
institutes, and acknowledge support from: FMSR (Austria); FNRS and FWO
(Belgium); CNPq, CAPES, FAPERJ, and FAPESP (Brazil); MES (Bulgaria);
CERN; CAS, MoST, and NSFC (China); COLCIENCIAS (Colombia); MSES
(Croatia); RPF (Cyprus); Academy of Sciences and NICPB (Estonia);
Academy of Finland, ME, and HIP (Finland); CEA and CNRS/IN2P3
(France); BMBF, DFG, and HGF (Germany); GSRT (Greece); OTKA and NKTH
(Hungary); DAE and DST (India); IPM (Iran); SFI (Ireland); INFN
(Italy); NRF and WCU (Korea); LAS (Lithuania); CINVESTAV, CONACYT,
SEP, and UASLP-FAI (Mexico); PAEC (Pakistan); SCSR (Poland); FCT
(Portugal); JINR (Armenia, Belarus, Georgia, Ukraine, Uzbekistan); MST
and MAE (Russia); MSTD (Serbia); MICINN and CPAN (Spain); Swiss
Funding Agencies (Switzerland); NSC (Taipei); TUBITAK and TAEK
(Turkey); STFC (United Kingdom); DOE and NSF (USA).

\bibliography{auto_generated}

\cleardoublepage \appendix\section{The CMS Collaboration \label{app:collab}}\begin{sloppypar}\hyphenpenalty=5000\widowpenalty=500\clubpenalty=5000\textbf{Yerevan Physics Institute,  Yerevan,  Armenia}\\*[0pt]
S.~Chatrchyan, V.~Khachatryan, A.M.~Sirunyan, A.~Tumasyan
\vskip\cmsinstskip
\textbf{Institut f\"{u}r Hochenergiephysik der OeAW,  Wien,  Austria}\\*[0pt]
W.~Adam, T.~Bergauer, M.~Dragicevic, J.~Er\"{o}, C.~Fabjan, M.~Friedl, R.~Fr\"{u}hwirth, V.M.~Ghete, J.~Hammer\cmsAuthorMark{1}, S.~H\"{a}nsel, M.~Hoch, N.~H\"{o}rmann, J.~Hrubec, M.~Jeitler, W.~Kiesenhofer, M.~Krammer, D.~Liko, I.~Mikulec, M.~Pernicka, B.~Rahbaran, H.~Rohringer, R.~Sch\"{o}fbeck, J.~Strauss, A.~Taurok, F.~Teischinger, C.~Trauner, P.~Wagner, W.~Waltenberger, G.~Walzel, E.~Widl, C.-E.~Wulz
% ...
\vskip\cmsinstskip
% ...
1:~~Also at CERN, European Organization for Nuclear Research, Geneva, Switzerland\\
% ...

\end{sloppypar}
\end{document}